# Deterministic and stochastic analysis of distributed order systems using operational matrix

Pham Luu Trung Duong and Moonyong Lee*

School of Chemical Engineering, Yeungnam University, Gyeongsan, Rep. of Korea
(Tel: 053-810-3241; Email: mynlee@yu.ac.kr)

*SUMMARY*

The fractional order system, which is described by the fractional order derivative and integral, has been studied in many engineering areas. Recently, the concept of fractional order has been generalized to the distributed order concept, which is a parallel connection of fractional order integrals and derivatives taken to the infinitesimal limit in delta order. On the other hand, there are very few numerical methods available for the analysis of distributed order systems, particularly under stochastic forcing. This paper first proposes a numerical scheme for analyzing the behavior of a SISO linear system with a single term distributed order differentiator/integrator using an operational matrix in the time domain under both deterministic and random forcing. To assess the stochastic distributed order system, the existing Monte-Carlo, polynomial chaos and frequency methods are first adopted to the stochastic distributed order system for comparison. The numerical examples demonstrate the accuracy and computational efficiency of the proposed method for analyzing stochastic distributed order systems.

KEY WORDS: Block pulse function, Distributed order system, Fractional calculus, Operational matrix, Stochastic process, Polynomial chaos

## 1. INTRODUCTION

Fractional / distributed order calculus has wide applicability across a range of disciplines, such as physics, biology, chemistry, finance, physiology and control engineering [1-6]. The distributed order (DO) equation, a generalized concept fractional order, was first proposed by M. Caputo in 1969 [7] and solved by him in 1995 [8]. The general solution of linear distributed order was discussed systematically in [9]. Later, the DO

*Correspondence to*: Moonyong Lee, School of Chemical Engineering, Yeungnam University, Gyeongsan, 712-749, Korea. E-mail: mynlee@yu.ac.kr,
Phone:+82 53 810 3241, Fax: +82 53 811 3262

concept was used to examine the diffusion equation [10], rheological properties of composite materials and other real complex physical phenomena [11-14]. Several different methods for time domain analysis of the DO systems have been reported [15-17]. On the other hand, the numerical method for an analysis of the DO operator is still immature and needs further development. In particular, there is no method to analyze distributed order systems under the excitation of random processes. This motivated the theme of this work: the development of a computational scheme for the analysis basic of a distributed order system with both deterministic and stochastic settings. The operational matrix (OP) has attracted considerable attention for the analysis of a range of dynamic systems [18-21]. The main characteristic of this technique is that different analysis problems can be reduced to a system of algebraic equations using different types of orthogonal functions, which greatly simplifies the problem [19]. On the other hand, to the author's knowledge, there are no reports on an analysis of distributed order systems using an operational matrix. Many natural systems often suffer stochastic noise that causes fluctuations in their behavior, making them deviate from deterministic models. Therefore, it is important to investigate such statistical characteristics of states (mean, variance of) for those stochastic systems. This problem is often called statistical analysis (or uncertainty quantification) of a system [22-24]. This paper proposes a numerical scheme based on the OP technique for the deterministic and statistical analysis of distributed order systems.

Monte-Carlo (MC) is a commonly used method for simulation of a stochastic model [25, 26]. The method relies on the sampling of independent realizations of random inputs according to their prescribed probability distribution. The data is fixed for each realization and the problem becomes deterministic. Solving the multiple deterministic realizations builds an ensemble of solutions, i.e. the realization of random solutions, from which statistical information can be extracted, e.g. mean and variance. Nevertheless, the method reveals slow convergence and large computational demand is typically needed. For example, mean values typically converge as $1/\sqrt{M}$, where $M$ is the number of samples.

Generalized polynomial chaos (gPC) [27-32] represents a more recent tool for quantifying the uncertainty within system models. The approach involves expressing stochastic quantities as the orthogonal polynomials of random input parameters. The method is actually a spectral representation in random space, and converges rapidly when the expanded function depends smoothly on random parameters. On the other hand, stochastic inputs of many systems involve random processes parameterized by truncated Karhunen-Loeve (KL) expansions, and the dimensionality of the KL expansions depends on the correlation lengths of these processes. For input processes with low correlation lengths, the number of dimensions required for an accurate representation can be extremely large.

The operational matrix method [29], where a system is described by a stochastic operator (operational matrix), is an alternative approach for simulation of stochastic integer order systems. This method involves the inverses of the stochastic operators as Neumann series, and is most effective for systems with inputs with low correlation lengths. On the other hand, it is restricted to small random parametric uncertainty.

In recent work [33], the authors introduced a hybrid spectral method, which combines the advantages of both the operational matrix and polynomial chaos, to simulate single input single output (SISO) stochastic fractional order systems. In the present study, the method reported in [33] is extended to statistical analysis of distributed order systems affected by stochastic fluctuations. Here, the stochastic operator is approximated using polynomial chaos instead of a Neumann series. This method provides the algebraic relationships between the first and second order stochastic moments of the input and output of a system, hence bypassing the KL expansions that can require large dimensions for accurate results. In contrast to the traditional operational matrix method, the proposed method is not limited by the magnitude of the uncertainty.

Section 2 briefly introduces a distributed order system and the operational matrix technique for uncertainty quantification in this system, leading to computation of the moments of random matrices. Section 3 summarizes the process of calculating the moments of the random matrices using a stochastic collocation. Section 4 provides examples to demonstrate the use of the proposed method. The results of the proposed deterministic system with distributed order were compared with those of other existing numerical and analytical methods. To assess the stochastic distributed order system, the MC, gPC and frequency methods were first adopted to the stochastic distributed order system for comparison because the analytical results were unavailable. The results from the proposed method were then compared with the numerical results from the MC, gPC and frequency methods.

## 2. PRELIMINARY OF FRACTIONAL AND DISTRIBUTED ORDER SYSTEM

### 2.1. Governing equation for system dynamics with fractional order dynamics

Fractional calculus considers the generalization of integration and differentiation operator to non-integer order [34-35]:

$$D_0^\alpha = \begin{cases} d^\alpha / dt^\alpha & \alpha > 0 \\ 1 & \alpha = 0 \\ \int_0^t (d\tau)^{-\alpha} & \alpha < 0 \end{cases} \qquad (1)$$

where $\alpha \in R$ is the order of the operator.

Among the many formulations of the generalized derivative, the Riemann-Liouville definition is used most often:

$$_{RL}D_0^\alpha f(t) = \frac{1}{\Gamma(m-\alpha)} (\frac{d}{dt})^m \int_0^t \frac{f(\tau)}{(t-\tau)^{1-(m-\alpha)}} d\tau \qquad (2)$$

where $\Gamma(x)$ denotes the gamma function; and $m$ is an integer satisfying $m-1 < \alpha < m$.

The Riemann-Liouville (RL) fractional integral of a function $f(t)$ is defined by the following:

$$_{RL}I_0^\alpha f(t) = \frac{1}{\Gamma(\alpha)} \int_0^t \frac{f(\tau)}{(t-\tau)^{1-\alpha}} d\tau \qquad (3)$$

Another popular definition of fractional order derivative is the Caputo (C) definition [36]:

$$_C D_t^\alpha = \frac{1}{\Gamma(m-\alpha)} \int_a^t (t-\tau)^{m-\alpha-1} f^{(m)}(\tau) d\tau \qquad (4)$$

The Laplace transform for a fractional order derivative under zero initial conditions can be defined as:

$$L\{D_0^\alpha f(t)\} = s^\alpha F(s) \qquad (5)$$

Note that under a zero initial condition, the two Riemann-Liouville and Caputo definitions are equivalent.

Therefore, a fractional order single input single output (SISO) system can be described by a fractional order differential equation:

$$a_0 D_0^{\alpha_0} y(t) + a_1 D_0^{\alpha_1} y(t) + ... + a_l D_0^{\alpha_l} y(t) = b_0 D_0^{\beta_0} u(t) + b_1 D_0^{\beta_1} u(t) + ... + b_m D_0^{\beta_m} u(t) \tag{6}$$

or by a transfer function:

$$G(s) = \frac{Y(s)}{U(s)} = \frac{b_m s^{\beta_m} + ... + b_0 s^{\beta_0}}{a_l s^{\alpha_l} + ... + a_0 s^{\alpha_0}} \tag{7}.$$

where $\alpha_i$ and $\beta_i$ are arbitrary real positive numbers; and $u(t)$ and $y(t)$ are the input and output of the system, respectively.

### 2.2. Distributed order systems

The distributed order differential operation is defined as follows [17]:

$$D_t^{\rho(\alpha)} f(t) = \int_{\gamma_1}^{\gamma_2} \rho(\alpha) D_t^{\alpha} f(t) d\alpha \tag{8}$$

where $\rho(\alpha)$ denotes the distribution function of order $\alpha$.

Therefore, the general form of the distributed order differential equation is:

$$\sum_{i=1}^{n} a_i D_t^{\rho_i(\alpha)} y(t) = \sum_{j=1}^{m} b_j D_t^{\rho_j(\alpha)} u(t) \tag{9}$$

For time domain analysis of the distributed-order, the integral in Equation (8) is discretized using the quadrature formula as follows [16,17]:

$$\int_{\gamma_1}^{\gamma_2} \rho(\alpha) D_t^{\alpha} f(t) d\alpha \approx \sum_{l=1}^{Q} \rho(\alpha_l)(D_t^{\alpha_l} f(t)) \upsilon_l \tag{10}$$

where $\alpha_l, \upsilon_l$ are the node and weight from the quadrature formula, respectively. In other words, the distributed order equation is approximated as a multi-term fractional order equation, and can be rearranged as Equation (6).

## 2.3. Operational matrices of block pulse function for the analysis of distributed order systems

Block pulse functions (BPFs) is a complete set of orthogonal functions and are defined over the time interval, $[0, \tau]$:

$$\psi_i = \begin{cases} 1 & \frac{i-1}{N}\tau \leq t \leq \frac{i}{N}\tau \\ 0 & \text{elsewhere} \end{cases} \tag{11}$$

where $N$ is the number of block pulse functions.

Therefore, any function that can be absolutely integrated on the time interval $[0, \tau]$ can be expanded into a series from the block pulse basis:

$$f(t) = \psi_N^T(t) C_f = \sum_{i=1}^{N} c_{f_i} \psi_i(t) \tag{12}$$

where $\psi_N^T(t) = [\psi_1(t), ..., \psi_N(t)]$ constitutes of the block pulse basis. From here, the subscript $N$ of $\psi_N^T(t)$ is dropped out for the convenience of notation.

The expansion coefficients (or spectral characteristics) can be evaluated as follows:

$$c_{f_i} = \frac{N}{\tau} \int_{[(i-1)/N]\tau}^{(i/N)\tau} f(t) \psi_i(t) dt \tag{13}$$

Furthermore, any function $g(t_1, t_2)$ absolutely integrable on the time interval $[0, \tau] \times [0, \tau]$ can be expanded as:

$$g(t_1, t_2) = \sum_{i=1}^{N} \sum_{j=1}^{N} c_{ij} \psi_i(t_1) \psi_j(t_2) = \psi^T(t_1) C_g \psi(t_2) \tag{14}$$

with expansion coefficients (or spectral characteristics) of:

$$c_{ij} = \left(\frac{N}{\tau}\right)^2 \int_{[(i-1)/N]\tau}^{(i/N)\tau} \int_{[(i-1)/N]\tau}^{(i/N)\tau} g(t_1, t_2) \psi_i(t_1) \psi_j(t_2) dt_1 dt_2 \tag{15}$$

Equation (3) can be expressed in terms of the operational matrix [18]:

$$I_0^\alpha f(t) = \psi(t)^T A_\alpha C_f \tag{16}$$

where the generalized operational matrix integration of the block pulse function, $A_\alpha$, is:

$$A_\alpha = P_\alpha^T = (\frac{\tau}{N})^\alpha \frac{1}{\Gamma(\alpha+2)} \begin{pmatrix} f_1 & f_2 & f_3 & \cdots & f_N \\ 0 & f_1 & f_2 & \cdots & f_{N-1} \\ \vdots & \ddots & \ddots & \ddots & \vdots \\ 0 & \cdots & \cdots & \cdots & f_1 \end{pmatrix}^T \qquad (17)$$

The elements of the generalized operational matrix integration can be given by:

$$f_1 = 1;\ f_p = p^{\alpha+1} - 2(p-1)^{\alpha+1} + (p-2)^{\alpha+1} \quad for\ p = 2,3... \qquad (18)$$

The generalized operational matrix of a derivative of order $\alpha$ is:

$$B_\alpha A_\alpha = I \qquad (19)$$

where *I* is the identity matrix.

The generalized operational matrix of derivative can be used to approximate Equation (2) as follows:

$$D_0^\alpha f(t) = \psi(t)^T B_\alpha C_f \qquad (20)$$

Therefore, using the operational matrix, the discretization of distributed order can be expressed as:

$$D_t^{\rho(\alpha)} f(t) = \int_{\gamma_1}^{\gamma_2} \rho(\alpha) D_t^\alpha f(t) d\alpha \approx \sum_{l=1}^{Q} \rho(\alpha_l)(D_t^{\alpha_l} f(t))\upsilon_l = \sum_{l=1}^{Q} \upsilon_l \rho(\alpha_l)(\psi(t)^T B_{\alpha_l} C_f) \qquad (21)$$

The distributed order system in Equation (9) can be rewritten in terms of the operational matrix, $A_G$:

$$A_G = [\sum_{i=1}^{n} a_i \sum_{l=1}^{Q} \upsilon_l \rho_i(\alpha_l) B_{\alpha_l}]^{-1} [\sum_{j=1}^{m} b_j \sum_{l=1}^{Q} \upsilon_l \rho_j(\alpha_l) B_{\alpha_l}] \qquad (22)$$

The input and output are related by the following equation:

$$C_Y = A_G C_U;\ Y(t) = (C_Y)^T \psi(t)\ ; U(t) = (C_U)^T \psi(t) \qquad (23)$$

The proposed method for the analysis of deterministic distributed order SISO system can be summarized in Algorithm 1.

Algorithm 1: Procedure for obtaining solution of deterministic distributed order Equation

a) Calculate the coefficients of expansions $C_U$ of the input in terms of block pulse function as in Equation (13).

b) Rewrite the distributed order differential Equation in Equation (9) with a suitable quadrature, in terms of operational matrix as in Equation (22). In this step, different quadrature rules can be used for the discretization of distributed order system such as Gauss-Legendre, Gauss Kronrod Patterson, etc.

c) The coefficients of the output is obtained as $C_Y = A_G C_U$.

d) The output is $Y(t) = (C_Y)^T \psi(t)$.

### 2.4. Stochastic analysis of distributed order systems

Consider the system described by Equation (9), which has its spectral characteristics of input and output linked by Equation (23). Assume that the system is excited by random forcing with given mean and covariance function as follows:

$$M_U(t) = E[U(t)] = (C_{m_U})^T \psi(t)$$
$$\kappa_{UU} = E\{[U(t_1) - M_U(t_1)][U(t_2) - M_U(t_2)]\} = \sum_{i=1}^{N}\sum_{j=1}^{N} \psi_i(t_1)\psi_j(t_2) c_{ij} = \psi(t_1)^T C_{K_{UU}} \psi(t_2) \quad (24)$$

where $E[]$ denote the expectation operator and the spectral characteristics of the mean and covariance function of the input are calculated in Equations (13) and (15).

Using the operational matrix, the spectral characteristics of the mean and covariance of the output are given by the following [33] (the details are available in the supplementary material):

$$C_{m_Y} = E[A_G] C_{m_U}$$
$$C_{K_{YY}} = E[A_G \{C_{K_{UU}} + (C_{m_U})(C_{m_U})^T\} A_G^T] - C_{m_Y}(C_{m_Y})^T \quad (25)$$

Therefore,

$$m_Y(t) = \psi(t_1)^T C_{m_Y} = \psi(t_1)^T E[A_G] C_{m_U}$$
$$\kappa_{YY}(t_1,t_2) = \psi(t_1)^T C_{\kappa_{YY}} \psi(t_2) = \psi(t_1)^T E[A_G \{C_{\kappa_{UU}} + (C_{m_U})(C_{m_U})^T\} A_G^T] \psi(t_2) - \psi(t_1)^T C_{m_Y} (C_{m_Y})^T \psi(t_2)$$
(26)

Random parameters $a_i, b_j$ result in the random operational matrix $A_G$ in Equations (25) and (26), and its moment can be estimated by a stochastic collocation method, which is described in the next section. When the parameters $a_i, b_j$ are deterministic, Equations (25) and (26) become:

$$m_Y(t) = \psi(t_1)^T A_G C_{m_U}$$
$$\kappa_{YY}(t_1,t_2) = \psi(t_1)^T A_G \{C_{\kappa_{UU}} + (C_{m_U})(C_{m_U})^T\} A_G^T \psi(t_2) - \psi(t_1)^T C_{m_Y} (C_{m_Y})^T \psi(t_2)$$
(27).

**Remarks**: The relationship in Equation (26) is invariant with respect to the orthogonal polynomial used to construct the operational matrix of the fractional order integral and derivative. The relationship in Equation (26) is only available for linear system.

### 3. STOCHASTIC COLLOCATION FOR OPERATIONAL MATRIX

A stochastic collocation method, which is described briefly below, is based on the gPC and can easily estimate the means and variances of complex dynamics. Hence, it has been used to estimate the moment of the random matrix in Equation (26).

- Assume that a random operational matrix has the form:

$$A = A(\xi) \tag{28}$$

where $\xi = (\xi_1, \xi_2, ..., \xi_n)$ is a vector of independent random parameters with probability density functions $\rho_i(\xi_i) : \Gamma_i \to R^+$. Vector $\xi$ has the joint probability density function of $\rho = \prod_{i=1}^{n} \rho_i$ with the support,

$$\Gamma \equiv \prod_{i=1}^{n} \Gamma_i \in R^{+n}.$$

- Choose a suitable quadrature set $\{\xi_i^{(m)}, w^{(m)}\}_{m=1}^{q_i}$ for each random parameter according to the probability density so that a one-dimensional integration can be approximated as accurate as possible

by:

$$\int_{\Gamma_i} A(\xi_i)\rho_i(\xi_i)d\xi_i = \sum_{i=1}^{q_i} A(\xi_i^{(m)})w_i^{(m)} \tag{29}$$

where $\xi_i^{(m)}$ is the $m^{\text{th}}$ node and $w^{(m)}$ is the corresponding weight.

- Construct a multi-dimensional cubature set by tensorization the one-dimensional quadrature set over all the combined multi-index $(j_1,\cdots,j_n)$. Because manipulation of the multi-index $(j_1,\ldots,j_n)$ is cumbersome in practice, a single index is preferable for manipulating these equations. The multi-index is often replaced by a graded lexicographic order index **j** [27]. Because the weighting functions of the cubature are the same as the probability density functions, the moment of the random matrix can be approximated by:

$$E[A] = \int_{\Gamma} A(\xi)\rho(\xi)d\xi = \sum_{\mathbf{j}=1}^{Q} A(\xi^{(\mathbf{j})})\mathbf{w}^{(\mathbf{j})} = \sum_{j_1=1}^{q_1}\cdots\sum_{j_n=1}^{q_n} A(\xi_1^{(j_1)},\ldots,\xi_n^{(j_n)})(w_1^{(j_1)}\ldots w_n^{(j_n)}) \tag{30}$$

The MATLAB suite, OPQ, can be used to obtain one-dimensional quadrature sets and their corresponding orthogonal polynomials (polynomial chaos) with respect to the different density weights [36].

The algorithm of the proposed method for the analysis of stochastic system can be summarized below in Algorithm 2.

For more clearly understanding of the algorithm, a similar algorithm is depicted graphically in [33] for the analysis of stochastic linear fractional order systems.

Algorithm 2: Procedure for obtaining solution of stochastic distributed order equation

a) Calculate the coefficients $C_{m_R}, C_{\kappa_{RR}}$ of expansions of the mean and covariance of the input as in Equations (13) and (15).

b) Rewrite the distributed order differential equation in (9) in terms of operational matrix as in Equation (22).

c) The coefficients of expansions of the mean and covariance function of the output are obtained from Equation (25). In Equation (25), the moments of several random matrices need to be calculated. A moment of a random matrix is calculated by the stochastic collocation method as in Equation (30).

d) Finally, the mean and covariance of the output are obtained as in Equation (26).

## 4. EXAMPLES

### 4.1. Example 1.

Consider a deterministic distributed order integrator:

$$\int_{0.5}^{0.8} s^{-\alpha} d\alpha \tag{31}$$

The analytical inverse Laplace transform of Equation (31), which is a system impulse response, can be expressed by the following [15]:

$$h(t) = \frac{1}{\pi} \int_{0}^{\infty} \frac{e^{-xt}}{(\ln(x))^2 + \pi^2} [x^{-0.5}(\sin(0.5\pi)\ln(x) + \pi\cos(0.5\pi)) - x^{-0.8}(\sin(0.8\pi)\ln(x) + \pi\cos(0.8\pi))]dx. \tag{32}$$

For the proposed method, the distributed order is discretized with the 3 points Gauss-Legendre quadrature rule obtained from the OPQ suite, and the operational matrix for this distributed integrator can be calculated by:

$$A_G = \sum_{i=1}^{3} A_{\alpha_i} \upsilon_i \tag{33}$$

where $\alpha_i, \upsilon_i$ are the nodes and weights of the quadrature, respectively.

Because the input here is a Dirac delta function, a rectangular pulse is used to approximate the singular Dirac delta functions in the proposed method. Therefore, the system output can be calculated as follows:

$$\begin{aligned} C_y &= A_G C_u; \\ y(t) &= \boldsymbol{\psi}^T(t) C_y \end{aligned} \tag{34},$$

where $y(t)$ is the system output and $C_u; C_y$ are the spectral characteristics of input and output, respectively.

Figure 1 presents the impulse response for this system obtained by Equation (32), the numerical inverse Laplace transform (NILT) [37], and the proposed method.

The NILT method is based on the application of fast Fourier transformation with quotient difference algorithm. Performances of the two methods were compared in terms of the absolute error in Figure 2. In the proposed method, the number of block pulse functions was chosen in such a way that the computational time of the two methods is approximately equivalent. As seen from Figure 2, the proposed method gave better performance in accuracy than the NILT method. In addition, the proposed method leads to algebraic equation which is more convenient to manipulate. When the system includes several simple components, the block algebra (similar to block algebra for transfer function) can be applied to obtain the system operational matrix from the operational matrix of components [33].

### 4.2. Example 2.

Consider an initial value problem for a distributed order relaxation equation taken from reference [17]:

$$\begin{aligned} &_0D_t^{\rho(\alpha)} y(t) + 0.1 y(t) = 0 \\ &y(0) = 1 \\ &\rho(\alpha) = 6\alpha(1-\alpha), \ 0 \le \alpha \le 1 \end{aligned} \tag{35}$$

Similar to matrix approach reported in [17], to use the operational matrix approach, Equation (35) needs to be converted to an equivalent problem with a zero initial condition as follows:

$$y(t) = x(t) + 1$$
$$_0D_t^{\rho(\alpha)}x(t) + 0.1x(t) = -0.1 \qquad (36)$$
$$x(0) = 0;$$

where $x(t)$ is an additional variable for converting the problem into a zero initial condition.

The solution of this equivalent problem can be approximated in term of the OP of the fractional order derivative $B_\alpha$ as follows:

$$C_x = (A_G + 0.1I)^{-1}C_{-0.1}; A_G = \sum_{i=1}^{3} B_{\alpha_i} \upsilon_i; x(t) = \psi^T(t)C_x \qquad (37)$$

where $C_{-0.1}$ is the spectral characteristics of -0.1.

Figure 3 shows the result of the computations using the proposed method and the matrix approach [17]. Similar to Example 1, the 3 points Gauss-Legendre quadrature is used to discretize the distributed order term in the proposed method. The values of the solution using the proposed method are in perfect agreement with the method reported in [17].

In the matrix approach [17], a fractional difference approximation is used for the approximated solution. In the proposed method, the solution is found in the form of series of orthogonal functions. Since there is no analytical solution for this example, it is not possible to compare the accuracy of the two methods. The main advantage of the proposed method is to provide a unified framework for treating both deterministic and stochastic SISO systems as will be shown in several next examples.

### 4.3. Example 3.

This example considers a statistical analysis of double $\delta$ function distributed order integrator taken from [38]:

$$D_t^{\rho(\alpha)} y(t) = u \qquad (38),$$

where $\rho(\alpha) = a_1\delta(\alpha - \alpha_1) + a_2\delta(\alpha - \alpha_2)$, $\delta()$ is Dirac-Delta function. The Equation (38) is actually a double fractional integrator:

$$a_1 D_0^{\alpha_1} y(t) + a_2 D_0^{\alpha_2} y(t) = u(t) \tag{39}$$

We consider the case where the input $u(t)$ is an ideal white noise with zero mean and covariance function

$$\kappa_{UU}(\tau) = \delta(\tau) = \delta(t_1 - t_2) \tag{40}$$

The variance of the output is given by [38]:

$$D_{y(t)} = \sigma^2(t) = \frac{1}{a_2^2} \int_0^t u^{2(\alpha_2 - 1)} \left[ \mathcal{E}_{\alpha_2 - \alpha_1, \alpha_2}\left(-\frac{a_1 u^{(\alpha_2 - \alpha_1)}}{a_2}\right) \right]^2 du \tag{41}$$

where $\mathcal{E}_{\alpha_2 - \alpha_1, \alpha_2}()$ is Mittag-Leffler function which can be compute by Matlab mlf.m function [39].

The operational matrix for system (39) is given by:

$$A_G = \sum_{i=1}^{2} B_{\alpha_i} a_i \tag{42}$$

This system has no random parameters. Therefore, the covariance function of system output can be approximated by Equation (27). The regularization technique is used to approximate the Dirac delta covariance function. The variance obtained by the proposed method is shown in Figure 4 for $a_1 = a_2 = 1; \alpha_1 = \frac{3}{4}$, and $\alpha_2 = 1$. The relative error of the proposed with respect to Equation (41) is also shown in this Figure 4. It can be seen that the result by the proposed method is quite satisfactory. Note that the infamous polynomial chaos method still has limitation for simulation white noise input which makes the proposed method become more attractive.

### 4.4. Example 4.

Consider an optimal damping system with a distributed order system in [17]:

$$G(s) = \frac{1}{s^2 + 10 \int_{0.8015}^{0.8893} s^\alpha d\alpha + 1} \tag{43}$$

under the excitation of ideal white noise with a covariance function $\delta(\tau) = \delta(t_1 - t_2)$ and zero mean. Again, this is a linear system and the mean of input is zero, the mean of the output is zero. This system has no random parameters. The covariance function of system output is approximated by Equation (27). The same regularization technique is used to approximate the Dirac delta covariance function, as shown in Example 1. The frequency method was used for comparison. Similar to integer order system, when the input is ideal white noise, the steady state variance of the output can be calculated using the frequency method as follows:

$$D_{yss} = \|G\|_2^2 = \frac{1}{2\pi j}\int_{-j\infty}^{j\infty} G(s)G(-s)ds \qquad (44)$$

where $\|\ \|_2$ denotes the $H_2$ norm of the system. Here, the $H_2$ norm of the distributed order system was calculated directly by a numerical integration of Equation (44) using the quadgk function of MATLAB. The main difference between the proposed and frequency methods is that the frequency method can only give exact solutions at steady state.

Figure 5 shows the variance of the system output using the proposed and frequency methods. The proposed method agrees well with the result by the frequency method, despite the fact that in the proposed method, the distributed order is discretized using a quadrature method.

**Remark:** For the frequency method, the steady state variance of the output (the $H_2$ norm) is evaluated directly without discretizing the distributed order term.

### 4.5. Example 5.

This case considers a distributed order with both random parameters and random forcing. The system is as follows:

$$G(s) = \frac{Y(s)}{U(s)} = \frac{1}{s^2 + a\int_{0.8015}^{0.8893} s^\alpha d\alpha + b} \qquad (45),$$

where $a$ is a uniform random variable in [9.5,10.5] and $b$ is uniform in [0.5,1]. The input $U(t)$ is a random process with a unit mean and covariance function of following:

$$\kappa_{UU}(t_1,t_2) = 0.25\text{sinc}(\frac{t_1 - t_2}{2\pi}) \qquad (46)$$

where the sinc function is defined as:

$$\text{sinc}(x) = \begin{cases} \sin(\pi x)/(\pi x) & \text{elsewhere} \\ 1 & \text{for } x = 0 \end{cases} \quad (47)$$

The mean and variance of the output calculated by the proposed method using Equation (26) were compared with the results from the polynomial chaos (gPC) and MC methods in Figure 6. In this study, the gPC and MC methods were first applied to the distributed order systems under stochastic forcing for comparison. In the MC and gPC methods, the distributed order term was discretized first and the routine fode_sol.m [40] was used to integrate the multi-term fractional order version of the distributed order system to obtain the results. The random process input $U(t)$ was parameterized using a non-canonical decomposition [41] (the details are available in the supplementary material). Table 1 lists the simulation parameters and computational times required for each method. Figure 6 and Table 1 show that the proposed method can provide similar accuracy with much less computational effort than the other methods. The advantage of the proposed method lies in its use of operational matrices: the mean and covariance of the output can be obtained directly from those of the input without any parameterization of the input.

The result of the gPC method was used as a reference solution for the MC and proposed methods. It is known that statistical errors exist when the MC method is used. In this study, 3 runs of the MC method (with 10000 samples for each run) proceeded. The absolute errors in the means and variances computed by the MC and proposed methods are shown in Figure 7. Note that the time domain based methods (proposed, MC, gPC) need to discretize the distributed order terms. Also, the gPC method still has a limitation for studying a system with white noise input. Figure 8 plots the means and variances computed by the gPC method with different number of quadrature nodes, and shows that the gPC method suffers from performance degradation over time. As time goes by, more cubature nodes have to be used for retaining the accuracy, which implies more computational load needed. This performance degradation is mainly due to the nonlinear term in the non-canonical decomposition used for the parameterization of random forcing. Figure 9 plots the means and variances by the proposed method for different number of block pulse functions. On the contrary to the gPC method, the proposed method provides consistent performance insensitive to the simulation parameters. The issues mentioned above also explain the choice of the frequency method as a reference solution in Example 4.

## 5. CONCLUSIONS

An operational matrix method was proposed to analyze distributed order systems in both a deterministic and stochastic setting. To analyze the system with a stochastic parameter perturbation, the stochastic collocation was used to estimate the random operator. This combines the advantages of both the operational matrix technique and polynomial chaos method. The use of operational matrices explicitly provides the relationship between the first and second order moment for the input and output of a system, bypassing parameterization of the random input when predicting the statistical characteristics and reducing the dimensions of the random space. This can also effectively handle a system with a low correlation length input (i.e. ideal white noise) by regularization. The numerical examples demonstrate the accuracy and computational efficiency of the proposed method for analyzing distributed order systems. However, the explicit relationship in Equation (26) is only available for linear system; the applicability of the proposed method is restricted to linear system only.

## ACKNOWLEDGEMENTS

This study was supported by Basic Science Research Program through the National Research Foundation of Korea(NRF) funded by the Ministry of Education, Science and Technology(2012012532).

Table 1. Simulation parameters and time profiles for obtaining the statistical characteristics by the MC, gPC, and proposed methods in Examples 4

| Example | Simulation parameters | | | Computational time (sec.) | | |
|---|---|---|---|---|---|---|
| | MC(Halton sampling) | gPC | Proposed | MC | gPC | Proposed |
| 4 | 10000 samples | 625 cubature nodes | 512 number of basis | 4773.9 | 614.0 | 95.5 |

**List of figures**

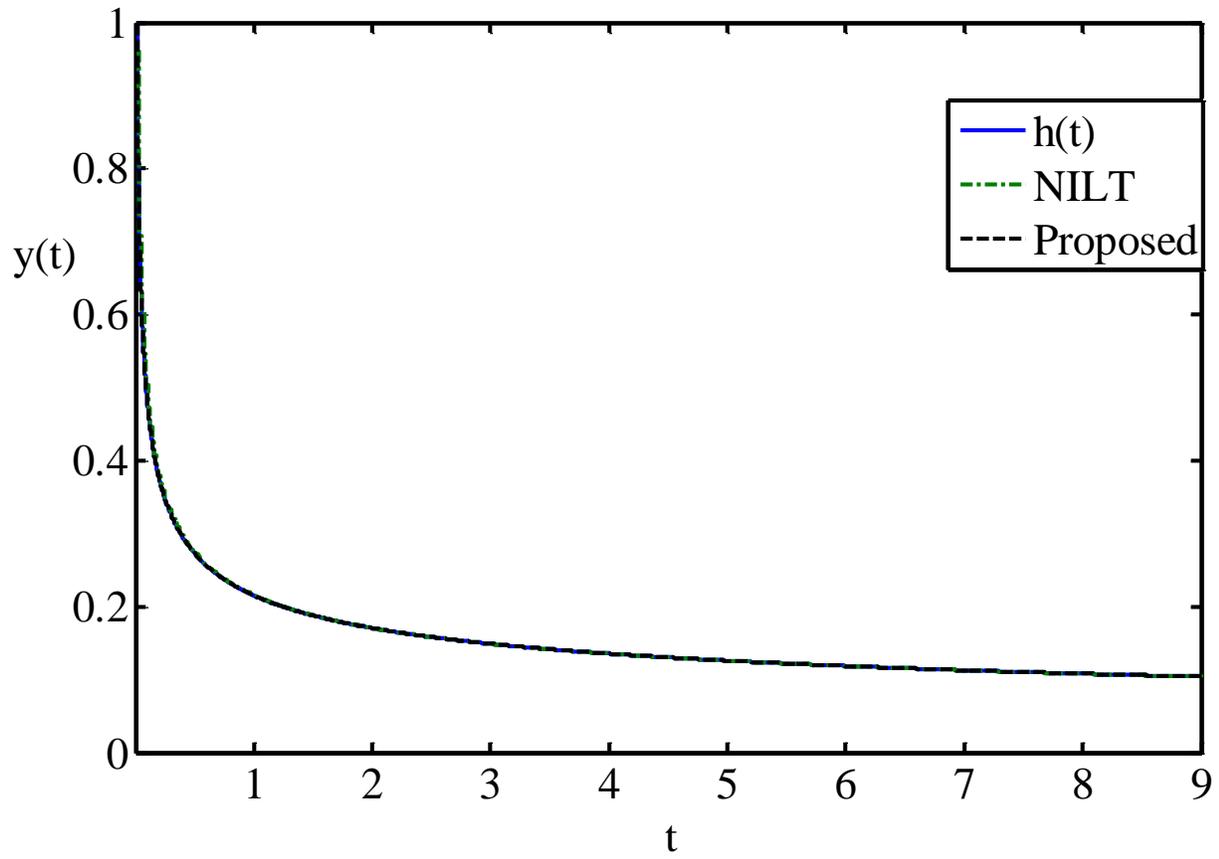

Figure 1. Impulse response of distributed order integrator.

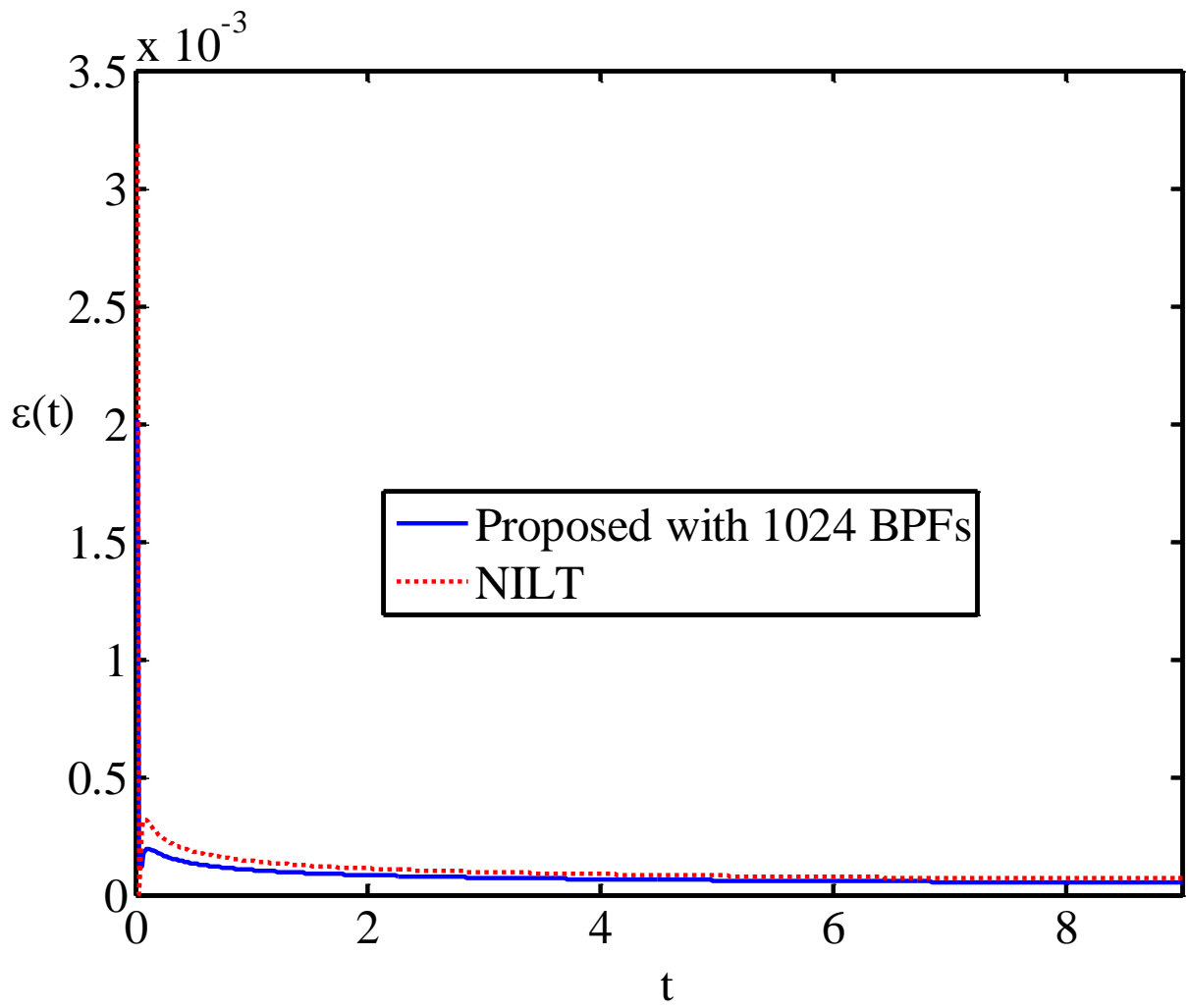

Figure 2. Absolute errors of the proposed and NILT methods.

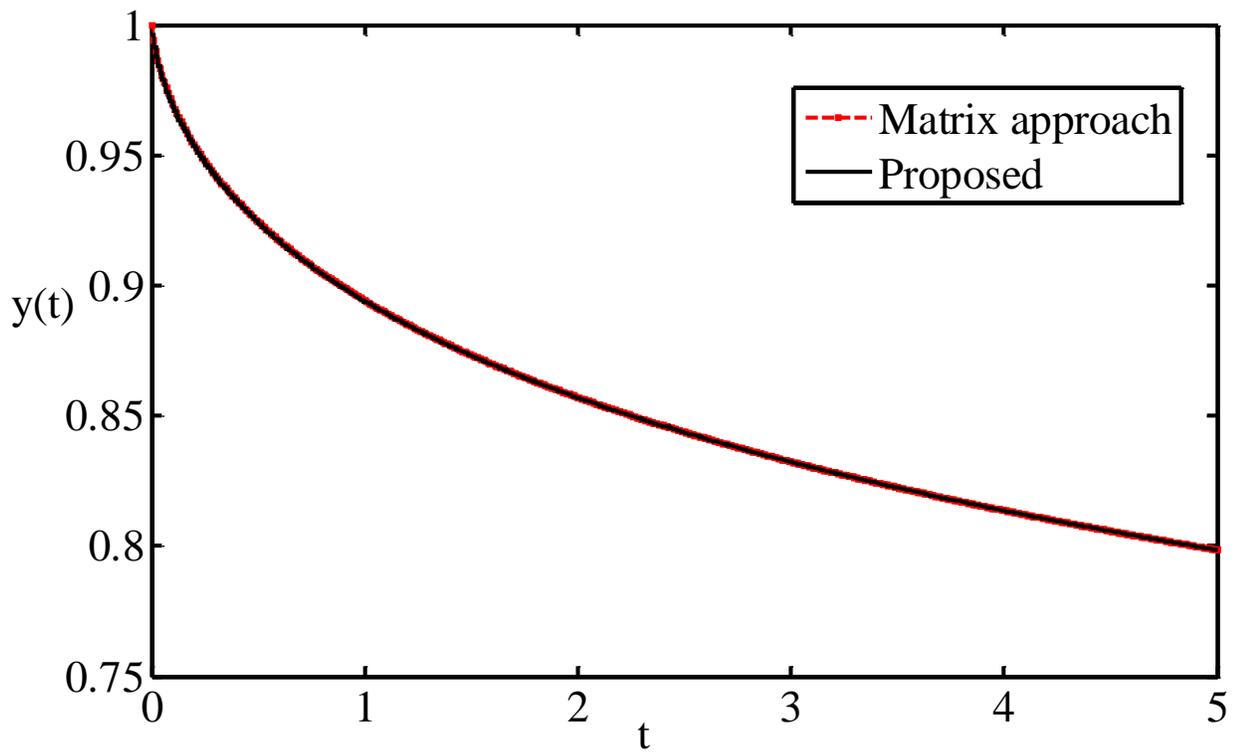

Figure 3. Solution of distributed order relaxation equation

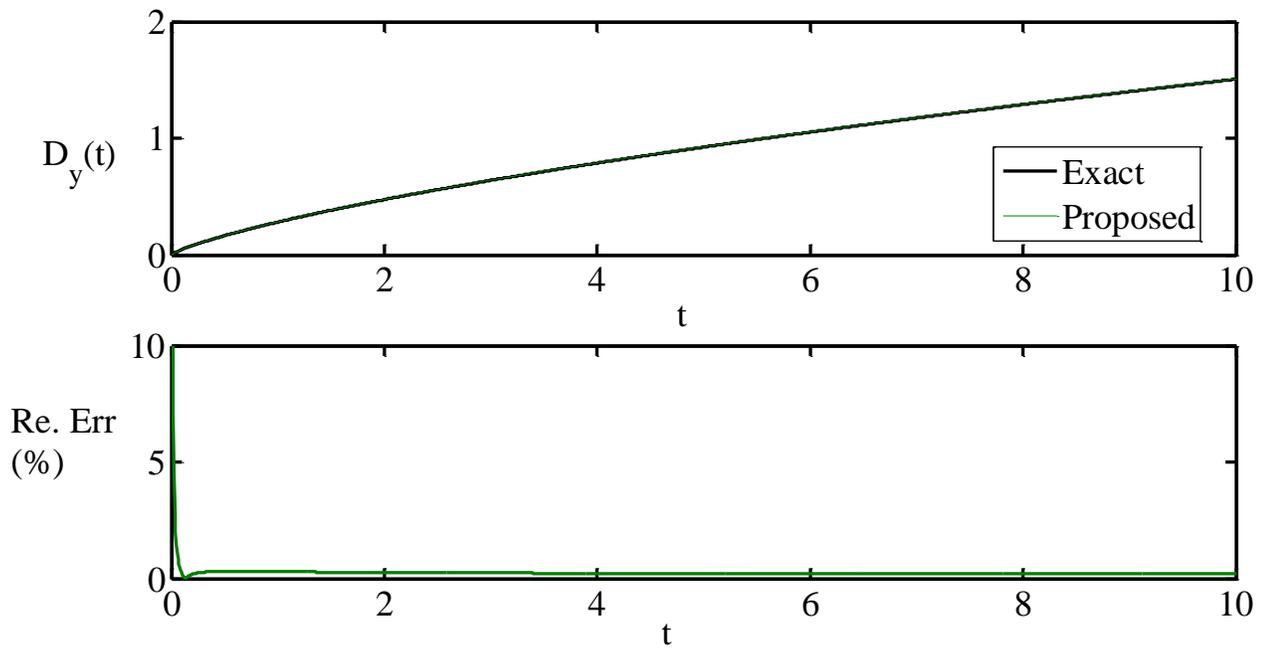

Figure 4.  Variances of the output in Example 3 with $a_1 = a_2 = 1; \alpha_1 = \dfrac{3}{4},$ and $\alpha_2 = 1$

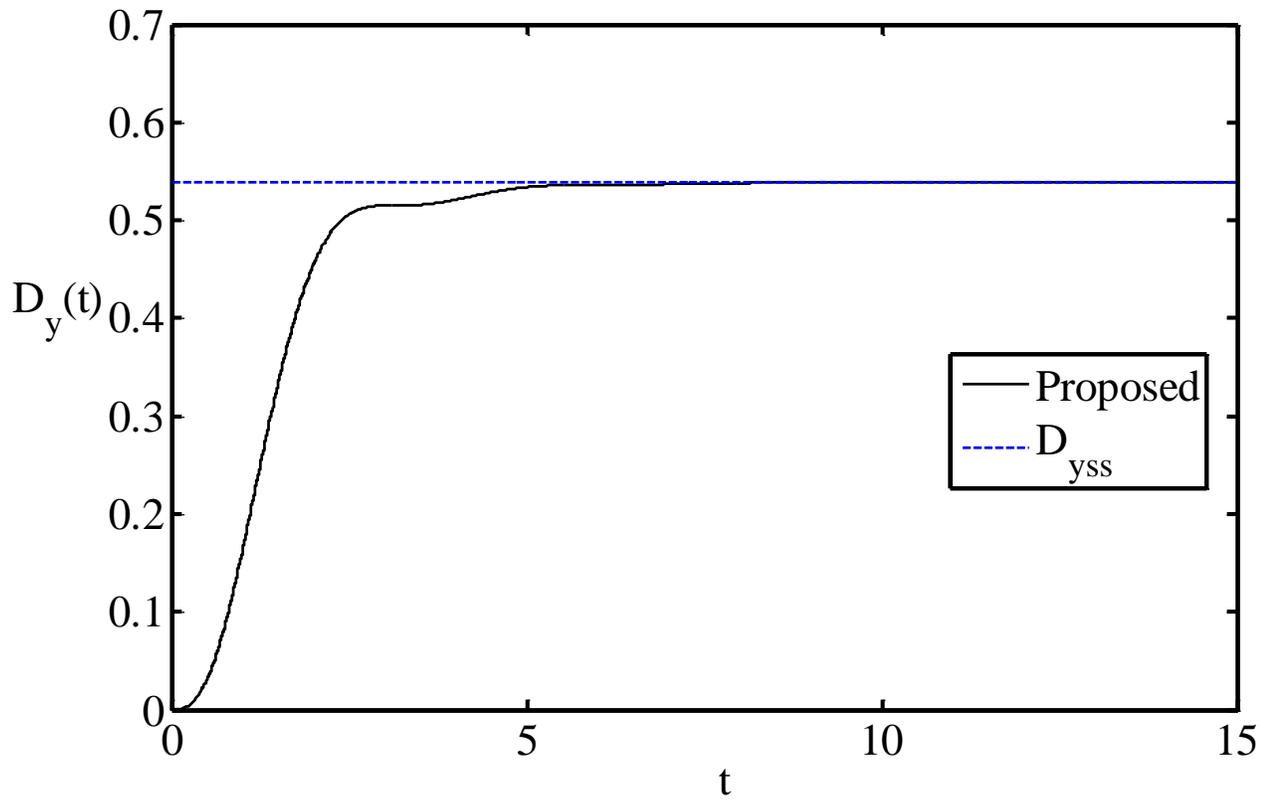

Figure 5. Variances of the output in Example 4

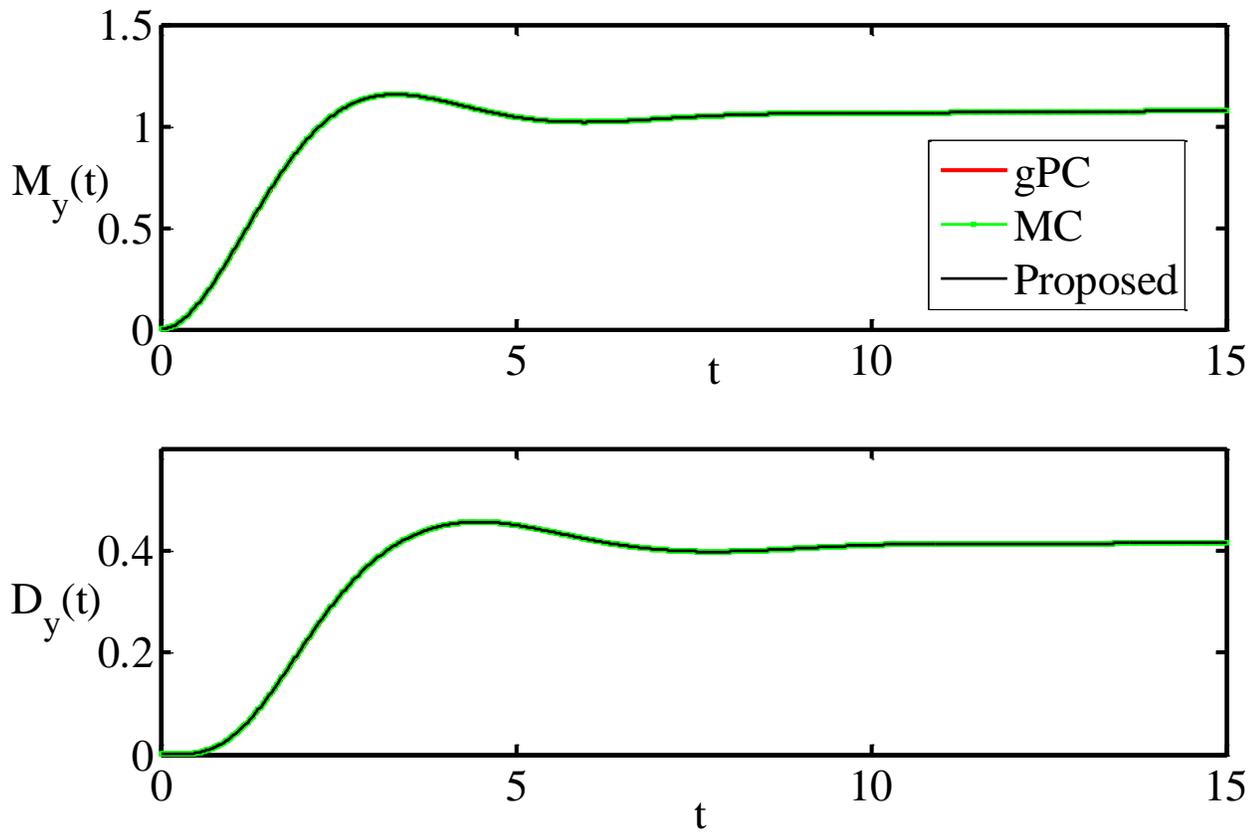

Figure 6. Mean and Variances of the output in Example 5

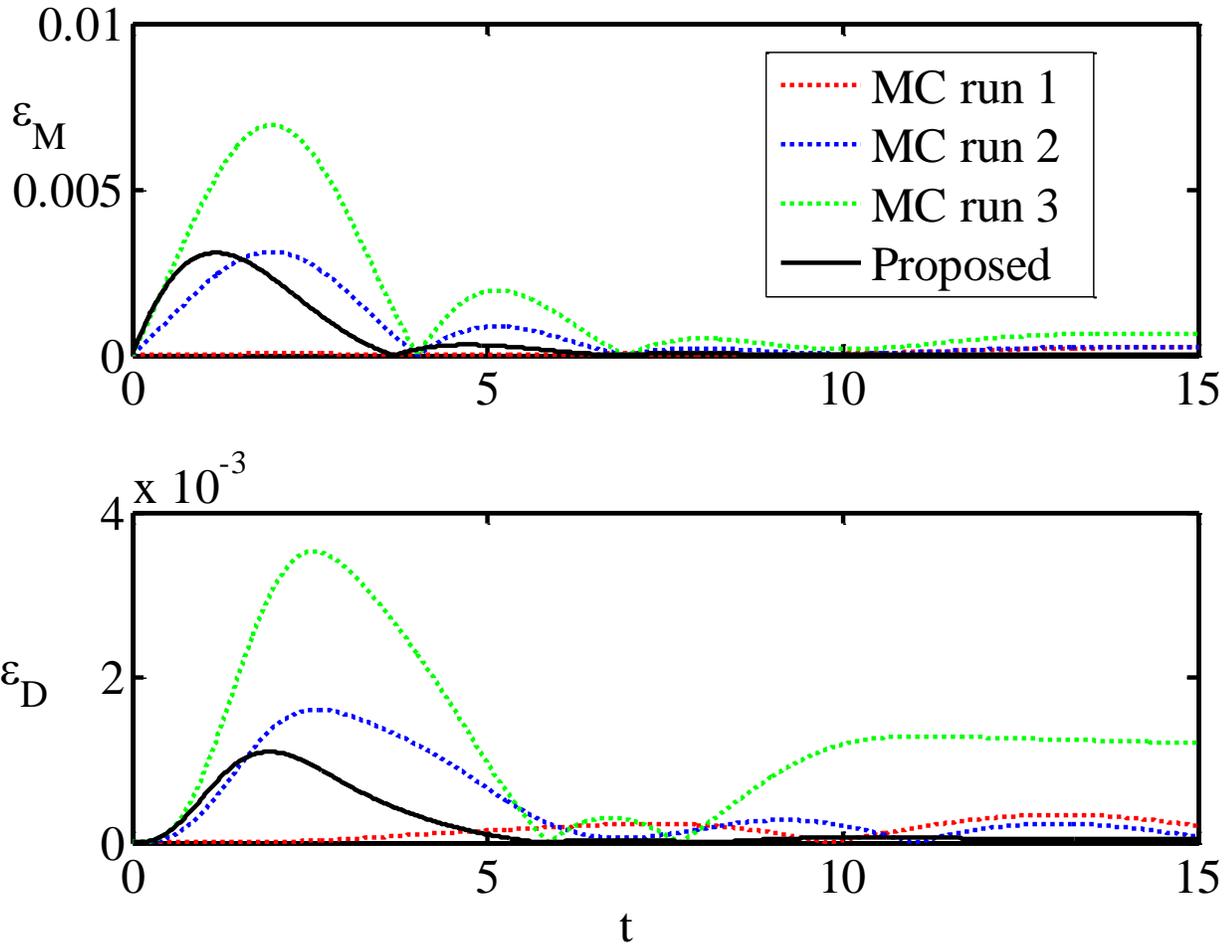

Figure 7. Absolute errors of the MC and proposed methods with respect to the gPC method for Example 5.

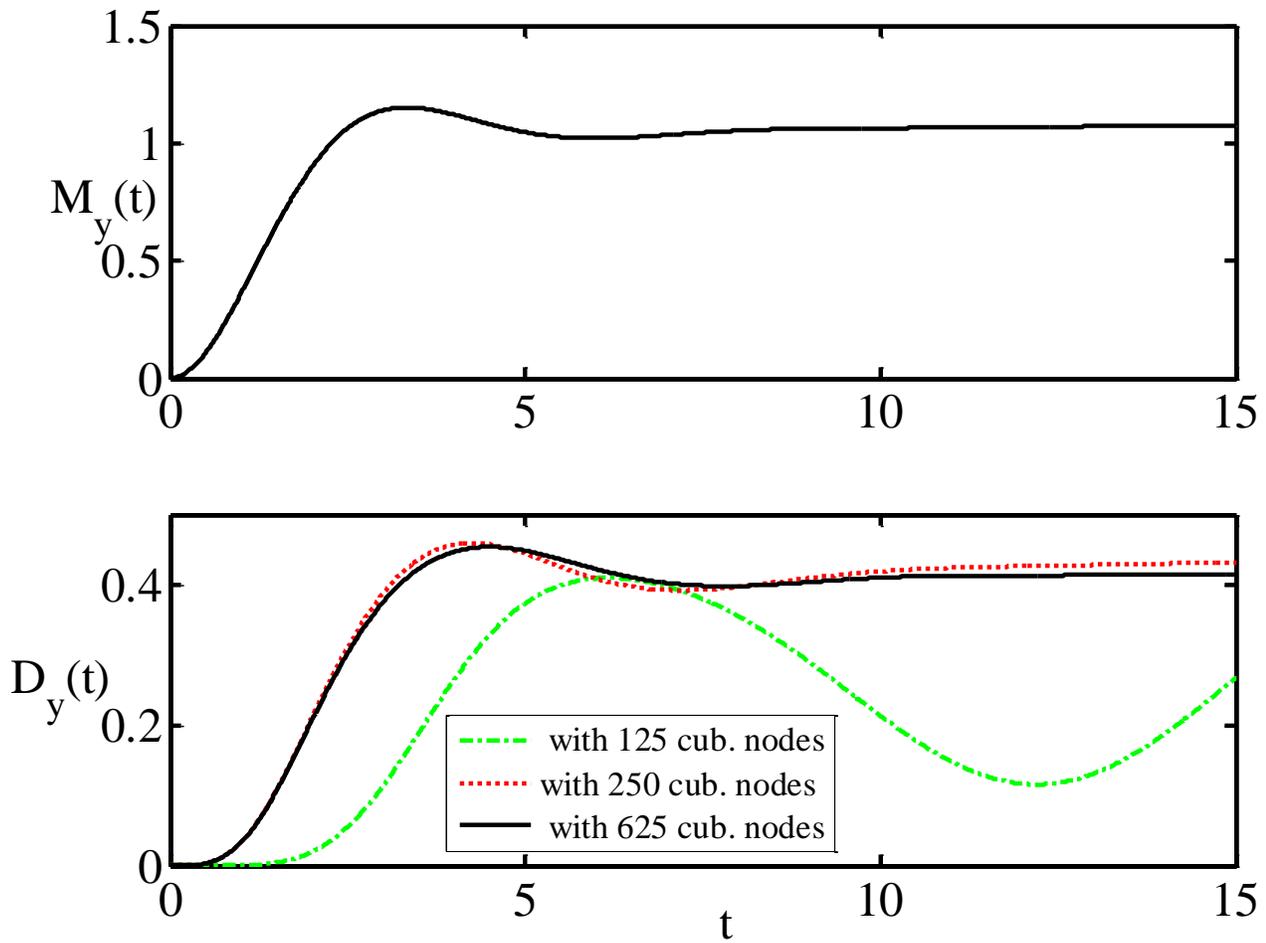

Figure 8. Mean and Variances of the output in Example 5 by the gPC method with different number of quadrature nodes

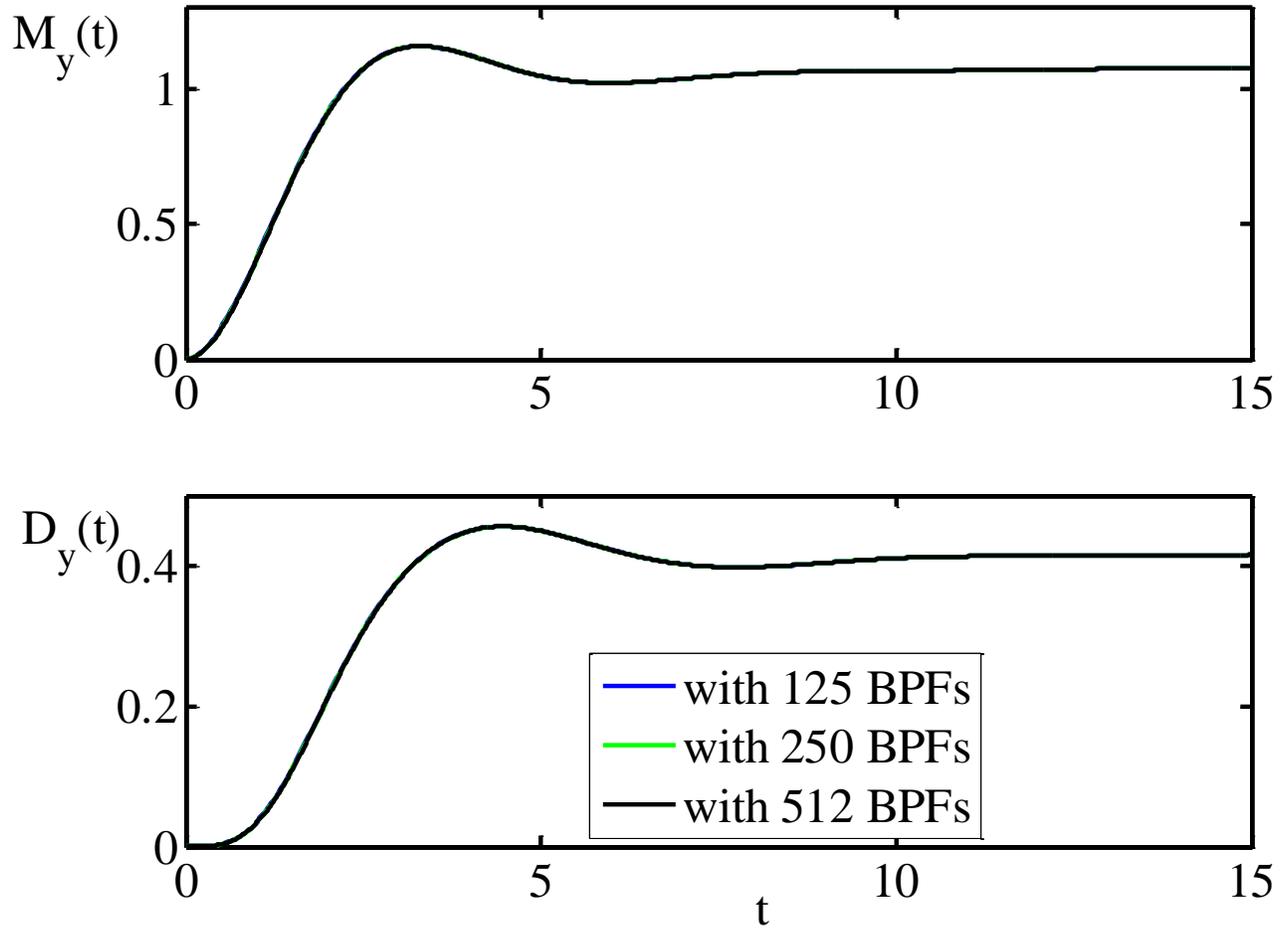

Figure 9. Mean and Variances of the output in Example 5 by the proposed method with different number of BPFs